\documentclass[]{pasj01}

\Received{}
\Accepted{}
 
\usepackage{url}
\usepackage{lineno}
 
\begin{document} 

\title{AGN number fraction in galaxy groups and clusters at {\boldmath $z <$} 1.4 from the Subaru Hyper Suprime-Cam survey}

 \author{
   Aoi			\textsc{Hashiguchi}		\altaffilmark{1},    
   Yoshiki		\textsc{Toba}			\altaffilmark{2,1,3,4$^\ast$\thanks{NAOJ Fellow}},
   Naomi		\textsc{Ota}			\altaffilmark{1},
   Masamune 	\textsc{Oguri}			\altaffilmark{5,6},     
   Nobuhiro 	\textsc{Okabe}			\altaffilmark{7,8,9},     
   Yoshihiro 	\textsc{Ueda}	   		\altaffilmark{10},  
   Masatoshi 	\textsc{Imanishi}		\altaffilmark{2}, 
   Satoshi 		\textsc{Yamada}			\altaffilmark{11},
   Tomotsugu 	\textsc{Goto}			\altaffilmark{12}, 
   Shuhei		\textsc{Koyama}			\altaffilmark{13}, 
   Kianhong 	\textsc{Lee}			\altaffilmark{14,2}, 
   Ikuyuki 		\textsc{Mitsuishi}		\altaffilmark{15}, 
   Tohru		\textsc{Nagao}			\altaffilmark{4},
   Atsushi J.	\textsc{Nishizawa}		\altaffilmark{16,17},   
   Akatoki 		\textsc{Noboriguchi}	\altaffilmark{18},
   Taira		\textsc{Oogi}			\altaffilmark{4},    
   Koki			\textsc{Sakuta}			\altaffilmark{15},    
   Malte		\textsc{Schramm}		\altaffilmark{19},     
   Mio			\textsc{Shibata}		\altaffilmark{1},  
   Yuichi		\textsc{Terashima}		\altaffilmark{20,4},
   Takuji		\textsc{Yamashita}		\altaffilmark{2},
   Anri			\textsc{Yanagawa}		\altaffilmark{1},  
   Anje			\textsc{Yoshimoto}		\altaffilmark{1}
}
\email{yoshiki.toba@nao.ac.jp}

\altaffiltext{1}{Department of Physics, Nara Women's University, Kitauoyanishi-machi, Nara, Nara 630-8506, Japan}
\altaffiltext{2}{National Astronomical Observatory of Japan, 2-21-1 Osawa, Mitaka, Tokyo 181-8588, Japan}
\altaffiltext{3}{Academia Sinica Institute of Astronomy and Astrophysics, 11F of Astronomy-Mathematics Building, AS/NTU, No.1, Section 4, Roosevelt Road, Taipei 10617, Taiwan}
\altaffiltext{4}{Research Center for Space and Cosmic Evolution, Ehime University, 2-5 Bunkyo-cho, Matsuyama, Ehime 790-8577, Japan}
\altaffiltext{5}{Center for Frontier Science, Chiba University, 1-33 Yayoi-cho, Inage-ku, Chiba 263-8522, Japan} 
\altaffiltext{6}{Department of Physics, Graduate School of Science, Chiba University, 1-33 Yayoi-Cho, Inage-Ku, Chiba 263-8522, Japan} 
\altaffiltext{7}{Department of Physical Science, Hiroshima University, 1-3-1 Kagamiyama, Higashi-Hiroshima,Hiroshima
739-8526, Japan}
\altaffiltext{8}{Hiroshima Astrophysical Science Center, Hiroshima University, 1-3-1 Kagamiyama, Higashi-Hiroshima,
Hiroshima 739-8526, Japan}
\altaffiltext{9}{Core Research for Energetic Universe, Hiroshima University, 1-3-1, Kagamiyama, Higashi-Hiroshima,
Hiroshima 739-8526, Japan}
\altaffiltext{10}{Department of Astronomy, Kyoto University, Kitashirakawa-Oiwake-cho, Sakyo-ku, Kyoto 606-8502, Japan}
\altaffiltext{11}{RIKEN Cluster for Pioneering Research, 2-1 Hirosawa, Wako, Saitama 351-0198, Japan}
\altaffiltext{12}{Institute of Astronomy, National Tsing Hua University, 101, Section 2. Kuang-Fu Road, Hsinchu 30013, Taiwan}
\altaffiltext{13}{Institute of Astronomy, Graduate School of Science, The University of Tokyo, 2-21-1 Osawa, Mitaka,Tokyo 181-0015, Japan}
\altaffiltext{14}{Astronomical Institute, Tohoku University, Aramaki, Aoba-ku, Sendai, 980-8578, Japan}
\altaffiltext{15}{Graduate School of Science, Division of Particle and Astrophysical Science, Nagoya University, Furocho, Chikusa-ku, Nagoya, Aichi 464-8602, Japan}
\altaffiltext{16}{DX Center, Gifu Shotoku Gakuen University, Takakuwa-Nishi, Yanaizucho, Gifu 501-6194, Japan}
\altaffiltext{17}{Institute for Advanced Research/Kobayashi Maskawa Institute, Nagoya University, Nagoya 464-8602, Japan}
\altaffiltext{18}{Center for General Education, Shinshu University, 3-1-1 Asahi, Matsumoto,
Nagano 390-8621, Japan}
\altaffiltext{19}{Universit\"at Potsdam, Karl-Liebknecht-Str. 24/25, D-14476 Potsdam, Germany}
\altaffiltext{20}{Department of Physics, Ehime University, 2-5 Bunkyo-cho, Matsuyama, Ehime 790-8577, Japan}











\maketitle

\begin{abstract}
One of the key questions on active galactic nuclei (AGN) in galaxy clusters is how AGN could affect the formation and evolution of member galaxies and galaxy clusters in the history of the Universe. 
To address this issue, we investigate the dependence of AGN number fraction ($f_{\rm AGN}$) on cluster redshift ($z_{\rm cl}$) and distance from the cluster center ($R/R_{\rm 200}$). 
We focus on more than 27,000 galaxy groups and clusters at $0.1 < z_{\rm cl} < 1.4$ with more than 1 million member galaxies selected from the Subaru Hyper Suprime-Cam.
By combining various AGN selection methods based on infrared (IR), radio, and X-ray data, we identify 2,688 AGN.
We find that (i) $f_{\rm AGN}$ increases with $z_{\rm cl}$ and (ii) $f_{\rm AGN}$ decreases with $R/R_{\rm 200}$.
The main contributors to the rapid increase of $f_{\rm AGN}$ towards high-$z$ and cluster center are IR- and radio-selected AGN, respectively.
Those results indicate that the emergence of the AGN population depends on the environment and redshift, and galaxy groups and clusters at high-$z$ play an important role in AGN evolution. 
We also find that cluster-cluster mergers may not drive AGN activity in at least the cluster center, while we have tentative evidence that cluster-cluster mergers would enhance AGN activity in the outskirts of (particularly massive) galaxy clusters.
\end{abstract}


\section{Introduction}
\label{Intro}
Galaxies are a key component in the formation of galaxy groups, clusters, and, eventually, large-scale structure in the Universe.
Given the fact that almost all galaxies have supermassive black holes (SMBHs) that could regulate the evolution of their host galaxies (e.g., \cite{Magorrian,Ferrarese,Woo}), active galactic nuclei (AGN) may play an important role even in the evolution of galaxy groups and clusters (see e.g., \cite{Fabian}, and references therein).
In other words, galaxy groups and clusters offer a unique laboratory for studying the relationship between AGN and the galaxies in which they reside.

To address how AGN could affect the formation and evolution of member galaxies and galaxy clusters in the history of the Universe, the AGN fraction ($f_{\rm AGN}$) of galaxy clusters is a crucial parameter.
$f_{\rm AGN}$ is often defined as the number of AGN among galaxy members in a cluster, and its redshift and environmental dependence are the main subject in this work.
The AGN fraction in galaxy groups and clusters and/or environmental dependence on AGN activity have been investigated by many authors based on infrared (IR) (e.g.,\cite{Krick,Tomczak,Santos,Somboonpanyakul}), radio (e.g., \cite{Best,Magliocchetti,Croston,Uchiyama}), and X-ray (e.g., \cite{Martini09,Pentericci,Ehlert14,Koulouridis}).
The AGN fraction has also been investigated through the semi-analytic galaxy formation model (e.g., \cite{Marshall}) (see also \cite{Munoz}).

For example, observational evidence has accumulated that the AGN fraction increases with redshift (e.g., \cite{Eastman,Martini09,Pentericci,Mishra,Bhargava}), as \citet{Butcher} discovered the redshift evolution of blue galaxies in galaxy clusters.
On the other hand, the environmental dependence of $f_{\rm AGN}$ has still been debated.
Some works reported that AGN favors a denser environment, i.e., AGN fraction in galaxy groups and clusters would be higher than that in field galaxies (e.g., \cite{Manzer}).
But \citet{Man} found little dependence of AGN fraction/activity on the environment (see also \cite{Santos}).
One of the reasons for this discrepancy would be the AGN sample selection.
Many works indeed relied on an AGN population selected with a single wavelength, which may induce an incomplete and biased AGN sample. 
However, multi-wavelength studies are still limited \citep{Galametz,Klesman12,Martini,Klesman14,Mo}.
Even those multi-wavelength studies often use $< 100$ clusters; they reported from a relatively small cluster sample that $f_{\rm AGN}$ increases with increasing redshift, and environmental dependence of AGN activity could depend on the wavelength used for AGN selection.
Nevertheless, since AGN is a relatively rare population, statistical uncertainties may not be negligible. 
A systematic investigation with a large number of clusters and a complete AGN sample is required to overcome this issue.

In this work, we investigate the redshift and environmental dependence of the AGN fraction in galaxy clusters in which we focus on large numbers of galaxy groups and clusters discovered by Hyper Suprime-Cam (HSC: \cite{Miyazaki}) Subaru Strategic Survey (HSC-SSP: \cite{Aihara18b,Aihara18a,Aihara19,Aihara22}).
The HSC-SSP has performed an optical imaging survey over about 1,200 deg$^2$ with five broadband filters and four narrowband filters (see \cite{Bosch,Coupon,Furusawa,Huang,Kawanomoto,Komiyama})\footnote{See also \citet{Schlafly,Tonry,Magnier,Chambers,Juric,Ivezic} for relevant papers.}.
This survey consists of three layers; Wide, Deep, and UltraDeep. 
This work utilizes s21a Wide-layer data obtained from 2014 March to 2021 January, providing forced photometry of $g$-, $r$-, $i$-, $z$-, and $y$-band with a 5$\sigma$ limiting magnitude of 26.8, 26.4, 26.4, 25.5, and 24.7, respectively.
For AGN identification among member galaxies, we combine several AGN selection methods with multi-wavelength data.
We construct mid-IR (MIR) AGN sources via IR color cuts while radio and X-ray AGN sources via luminosity cuts following \citet{Galametz}, which provides an unbiased AGN sample.

This paper is structured as follows. 
In section \ref{S_data_ana}, we describe the sample selection of galaxy clusters, member galaxies, and AGN.
The resultant dependence of AGN fraction on cluster redshift and distance from cluster center is presented in section \ref{Results}.
We then discuss possible uncertainty and selection bias of the results and revisit them from a multi-wavelength point of view in section \ref{S_dis}.
We also discuss how cluster-cluster mergers would enhance AGN activity.
The main conclusions are summarized in section \ref{S_summary}.
Throughout this paper, the adopted cosmology is a flat Universe with $H_0$ = 70 km s$^{-1}$ Mpc$^{-1}$, $\Omega_{\rm M}$ = 0.28, and $\Omega_{\rm \Lambda}$ = 0.72, which are the same as those adopted in \citet{Oguri18}.

\section{Data and analysis}
\label{S_data_ana}

\subsection{Sample selection}
\label{S_sample}
\begin{figure}[h]
 \begin{center}
 \includegraphics[width=0.45\textwidth]{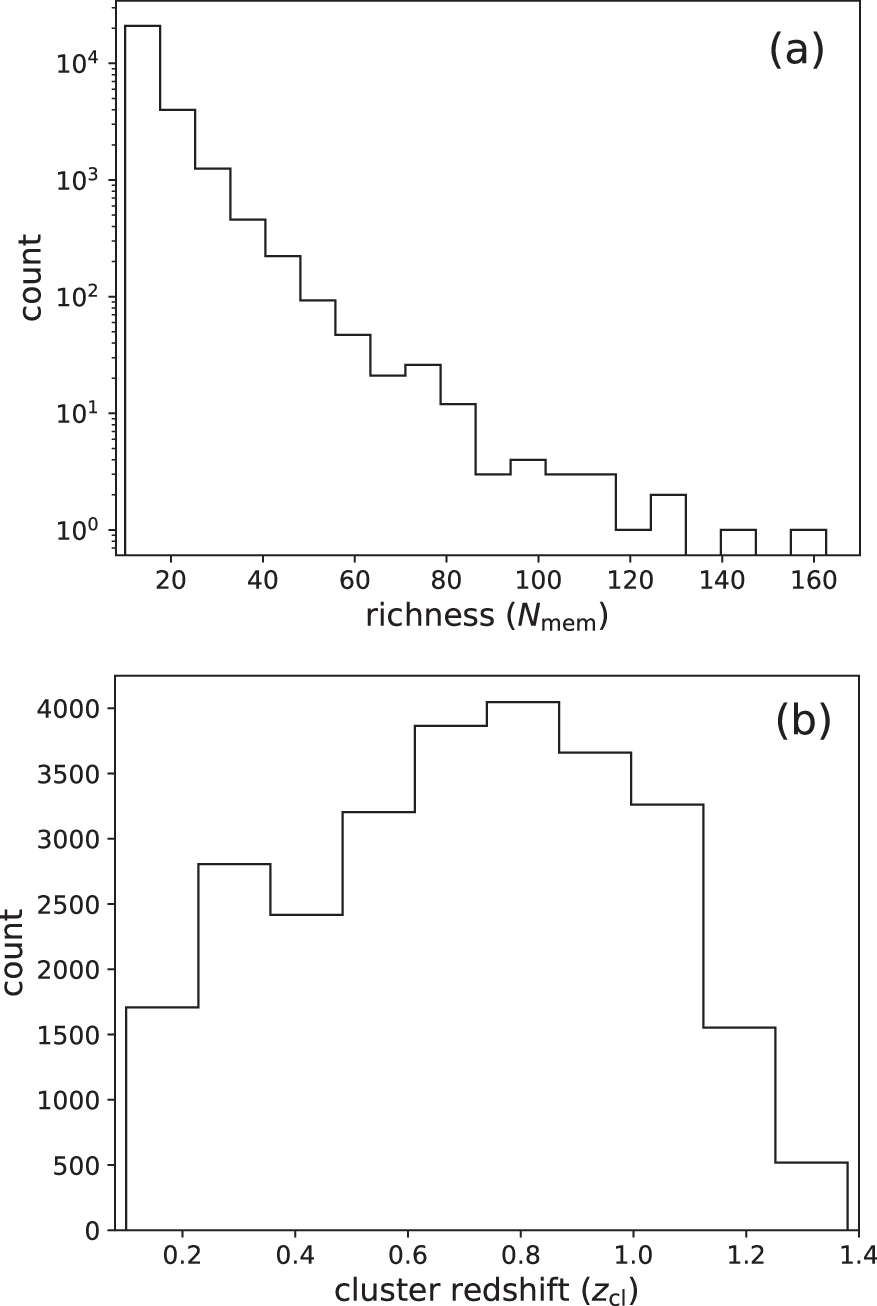}
 \end{center}
\caption{The distributions of (a) richness ($N_{\rm mem}$) and (b) cluster redshift ($z_{\rm cl}$) of CAMIRA clusters.}
\label{camira}
\end{figure}

We use an HSC-selected galaxy group and cluster catalog with the CAMIRA (Cluster-finding Algorithm based on Multi-band Identification of Red-sequence galaxies) algorithm \citep{Oguri} being applied to the HSC-SSP data (see \cite{Oguri18} for more detail\footnote{They essentially used $r$, $i$, and $z$ colors for HSC sources with $z_{\rm AB}$ $<$ 24.0.}).
We utilize the latest version ({\tt s21a\_v1}) of the CAMIRA catalog with bright star masks, which provides 27,037 galaxy groups and clusters with a richness of $N_{\rm mem} > 10$ in $\sim$ 1,027 deg$^2$.
The cluster redshift ($z_{\rm cl}$) is calculated in the process of CAMIRA cluster finding algorithm in which photometric redshifts of high-confidence cluster members, including brightest cluster galaxies (BCGs) are refined by maximizing a likelihood (see \cite{Oguri}).
The resultant $z_{\rm cl}$ of the CAMIRA groups/clusters distributes $0.1 < z_{\rm cl} < 1.4$ and is well determined with a bias of $\delta_{\rm z}$ of $\sim -0.001$, a scatter of $\sigma_{\rm z} \sim 0.008$ and an outlier rate of $f_{\rm out} \sim 0.02$ \citep{Oguri18}\footnote{\citet{Oguri18} evaluate the accuracy of $z_{\rm cl}$ from residual, ($z_{\rm cl} - z_{\rm BCG,spec}) / (1+z_{\rm BCG,spec})$. They define the bias ($\delta_{\rm z}$) and scatter ($\sigma_{\rm z}$) by the mean and standard deviation of the residual with 4$\sigma$ clipping. They then define the outlier fraction ($f_{\rm out}$) as the fraction of galaxies removed by the 4$\sigma$ clipping.}.
Figure \ref{camira} shows the distributions of $N_{\rm mem}$ and $z_{\rm cl}$ of CAMIRA clusters.
The CAMIRA catalog also contains 1,052,529 member galaxies, which serves as one of the largest member galaxy catalogs to date.

\subsection{Redshifts of member galaxies in galaxy clusters}
\label{s_member}

To select AGN based on the luminosities (see sections \ref{Radio_AGN} and \ref{X_AGN}) and obtain their physical properties, redshift information is required.
We first compile spectroscopic redshift ($z_{\rm spec}$) from literature such as the Sloan Digital Sky Survey (SDSS: \cite{York}) Data Release (DR) 15 \citep{Aguado} and Galaxy And Mass Assembly (GAMA: \cite{Driver}) DR2 \citep{Baldry}
(see also \cite{Tanaka,Nishizawa}).
For member galaxies without $z_{\rm spec}$, we employ the Direct Empirical Photometric code ({\tt DEmP}: \cite{Hsieh})\footnote{Other photometric redshift codes are also available provided by the HSC-SSP team (see \cite{Tanaka,Nishizawa} for more detail).}.
{\tt DEmP} is an empirical quadratic polynomial photometric redshift ($z_{\rm phot}$) fitting code demonstrating nice performance to red sequence galaxies (see \cite{Hsieh,Hsieh05} for a full description of this redshift code).
{\tt DEmP} also outputs physical properties of galaxies such as stellar mass ($M_*$) and star formation rate (SFR) that are widely used in previous works on CAMIRA clusters and/or environmental properties around AGN (e.g., \cite{Lin17,Jian,Shirasaki}).

\begin{figure}[h]
 \begin{center}
 \includegraphics[width=0.45\textwidth]{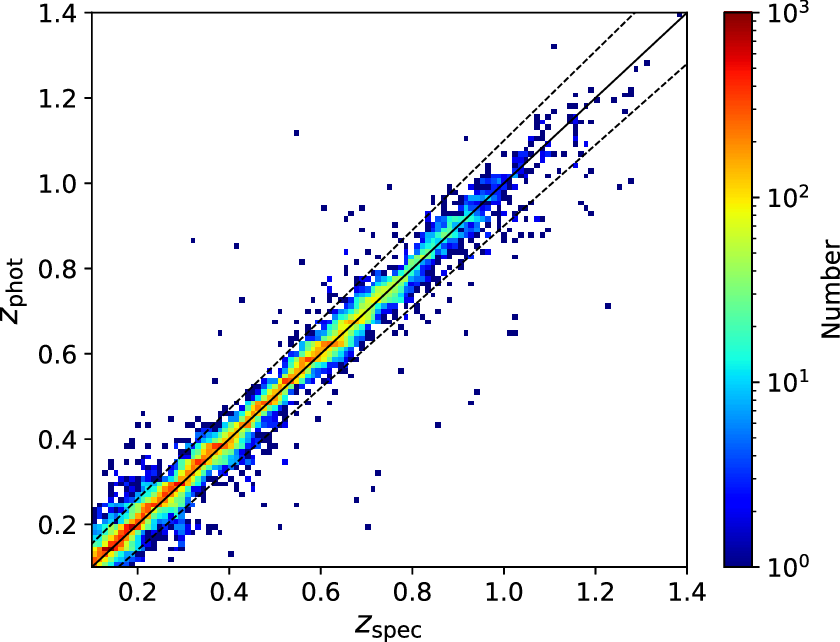}
 \end{center}
\caption{Comparison between $z_{\rm spec}$ and $z_{\rm phot}$ for member galaxies in CAMIRA clusters, color-coded by the number of objects per pixel. The black solid line represents $z_{\rm spec}$ = $z_{\rm phot}$. The black dotted lines denote $\Delta z/(1+z_{\rm spec)} = \pm 0.066$ corresponds to the 4$\sigma$ separation from mean residual.}
\label{DEmP}
\end{figure}

Figure \ref{DEmP} shows a comparison between $z_{\rm spec}$ and $z_{\rm phot}$ for member galaxies in CAMIRA clusters,
where 39,813 member galaxies with $z_{\rm spec}$ are plotted.
Following \citet{Oguri18}, we evaluate $\delta_{\rm z}$, $\sigma_{\rm z}$, and $f_{\rm out}$ of our sample.
The resulting values are $\delta_{\rm z}$ = 0.004, $\sigma_{\rm z}$ = 0.012, and $f_{\rm out}$ = 0.008, which indicates that $z_{\rm phot}$ of member galaxies is also well-determined as good accuracy as those for the cluster redshift.
Given this high quality, we do not consider the uncertainty of $z_{\rm phot}$ in this work, which has a negligible impact on the final results.
If a member galaxy has $z_{\rm spec}$, we use it.
Otherwise, we use $z_{\rm phot}$.
Hereafter, the redshift calculated in this way will be referred to as $z_{\rm mem}$.
Following \citet{Ando}, we narrow down the member galaxy sample to sources with $|z_{\rm cl} - z_{\rm mem}| \leq 0.05 \times ( 1 + z_{\rm cl})$ to pick up reliable member galaxies (see also section \ref{s_contami}).
As a result, 877,642 member galaxies in CAMIRA clusters are left, which is used for subsequent analysis.

\subsection{AGN identification in CAMIRA member galaxies}
\label{s_AGN}

To identify member galaxies hosting AGN in CAMIRA clusters, multi-wavelength catalogs with IR, radio, and X-ray are used.
For IR-selected AGN, we utilize a dedicated AGN catalog and select member galaxies with MIR-AGN by cross-matching it with CAMIRA member galaxies (section \ref{MIR_AGN}).
For radio and X-ray selected AGN, we first cross-match radio and X-ray catalogs with CAMIRA member galaxies and then extract radio and X-ray AGN by adopting luminosity cuts (sections \ref{Radio_AGN} and \ref{X_AGN}).
The basic information of each catalog and AGN selection criterion are summarized in table \ref{AGN_catalog}.
We note that it is hard to conduct a classical AGN selection with optical emission line diagnostics (e.g., \cite{BPT}) because spectroscopic completeness for CAMIRA member galaxies is quite low (38,319/877,642 $\sim$4.4\%).
Hence, we focus on AGN selections with IR, radio, and X-ray.

\begin{table*}[h]
\tbl{AGN sample used for this work.}{
\scalebox{0.85}[0.9]{
\begin{tabular}{crrcr}
\hline\hline
\multicolumn{1}{c}{Wavelength}	&  \multicolumn{1}{c}{Number of cataloged objects}	& \multicolumn{1}{c}{Reference}		&	\multicolumn{1}{c}{Selection criterion}  & \multicolumn{1}{c}{Number of selected AGN in CAMIRA clusters}	\\
\hline
  Mid-IR	&	4,543,530	&	\citet{Assef}	&	WISE color selections								&	  886  \\
  Radio 	& 	  946,432 	& 	\citet{Helfand} &	$L_{\rm 1.4 GHz} \geq 10^{24}$ W Hz$^{-1}$			&	1,588  \\
  X-ray 	& 	  602,543 	&	 \citet{Webb}	&	$L_{\rm X}$ (2--10 keV) $\geq 10^{42}$ erg s$^{-1}$ &     236  \\ \cline{5-5}
			&				&					&						& Total (after removing duplication): 	 2,688 \\
\hline
\end{tabular}
}
}
\label{AGN_catalog}
\end{table*}

\begin{figure*}[h]
 \begin{center}
 \includegraphics[width=\textwidth]{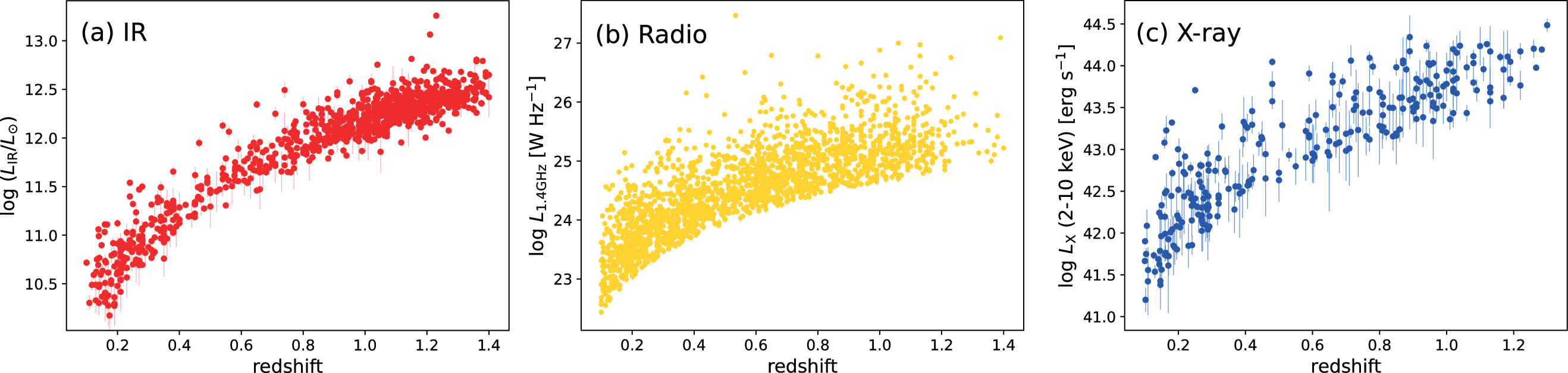}
 \end{center}
\caption{(a) IR luminosity ($L_{\rm IR}$), (b) radio luminosity at 1.4 GHz ($L_{\rm 1.4 GHz}$), and (c) hard X-ray luminosity in the 2--10 keV ($L_{\rm X}$ (2--10 keV)) as a function of redshift.}
\label{z_L}
\end{figure*}

\subsubsection{Mid-IR selected AGN}
\label{MIR_AGN}

The MIR selected AGN sample is constructed based on Wide-field Infrared Survey Explorer (WISE: \cite{Wright}). 
WISE performed an all-sky survey in four bands; W1 (3.6 $\micron$), W2 (4.5 $\micron$, W3 (12 $\micron$), and W4 (22 $\micron$), with an angular resolution of $\sim$6 arcsec.
In this work, we utilize the WISE AGN catalog provided by \citet{Assef}, who applied W1 -- W2  cut with a W2 magnitude dependence to the ALLWISE catalog\footnote{\url{https://wise2.ipac.caltech.edu/docs/release/allwise/}} (see \cite{Assef} for further details on the AGN selection).
We use the ``R90''  catalog, which consists of 4,543,530 robust AGN candidates with 90\% reliability across 30,093 deg$^2$ of the extragalactic sky.
There are 273,908 MIR-AGN in the survey footprint of the HSC--Wide region.

We then cross-match the MIR-AGN sample with CAMIRA member galaxies (i.e., HSC coordinates), using 3 arcsec as a search radius following \citet{Toba}.
As a result, 886 MIR-AGN are identified.
Among them, 92 WISE AGN have more than two HSC candidate counterparts.
In this work, we select the closest object as the counterpart.
We also visually check the WISE images for those MIR-AGN and confirm there are no spurious sources.
Figure \ref{z_L}(a) shows IR luminosity, $L_{\rm IR}$ (8-1000 $\mu$m), of MIR-AGN as a function of redshift, where $L_{\rm IR}$ is converted from monochromatic luminosity at 22 $\mu$m by using an empirical relation reported in \citet{Toba17b}.

\subsubsection{Radio selected AGN}
\label{Radio_AGN}

The radio-selected AGN sample is constructed based on the Very Large Array (VLA) Faint Images of the Radio Sky at Twenty-Centimeters (FIRST: \cite{Becker}), which realize a complete radio survey down to 1 mJy at 1.4 GHz over 10,000 deg$^2$.
The typical full width at half maximum (FWHM) of the beam is about 6 arcsec.
We utilize the final release catalog of FIRST \citep{Helfand}, which contains 946,432 radio sources.
There are 137,094 FIRST objects in the HSC--Wide region.
Following \citet{Yamashita}, we further constrain the sample with $F_{\rm int} > 1$ mJy and $P(S) < 0.05$ where  $F_{\rm int}$ and $P(S)$ is the integrated flux density at 1.4 GHz and a probability that a source is a spurious detection near a bright source, respectively.
The above cuts provide 96,605 clean radio sources.

We then cross-match the radio sample with CAMIRA member galaxies using a 1 arcsec search radius following \citet{Yamashita}, which leaves 2,084 radio sources in CAMIRA clusters.
Among them, 39 FIRST radio sources have more than two candidate counterparts of HSC sources.
We choose the closest object as the counterpart.
We then estimate their rest-frame 1.4 GHz radio luminosity ($L_{\rm 1.4 GHz}$) by assuming a power-law radio spectrum index of $\alpha_{\rm radio} = -0.7$ (e.g., \cite{Condon,Yamashita}).
Figure \ref{z_L}(b) shows resultant $L_{\rm 1.4 GHz}$ of 2,084 radio sources as a function of redshift.
Finally, we extract radio-bright sources with $L_{\rm 1.4 GHz} \geq 1.0 \times 10^{24}$ W Hz$^{-1}$ as AGN in the same manner as e.g., \citet{Tadhunter,Yamashita}.
As a result, 1,588 radio-AGN are identified.
We also visually check the FIRST images for those radio-AGN and confirm there are no spurious sources.
Note that radio luminosity cut is not the only way to select radio AGN.
We will discuss consistency with another selection method in section \ref{dis_Radio}.

\subsubsection{X-ray selected AGN}
\label{X_AGN}

The X-ray-selected AGN sample is constructed based on the XMM-Newton serendipitous source catalog (4XMM-DR11: \cite{Webb}).
This catalog contains 602,543 unique X-ray sources\footnote{\url{http://xmmssc.irap.omp.eu/Catalogue/4XMM-DR11/4XMM_DR11.html}}, which realize a complete X-ray survey down to $\sim 10^{-14}$ erg s$^{-1}$ cm$^{-2}$ in the total photon-energy band (0.2 --12 keV) over 1,200 deg$^2$.
The FWHM of the point spread function (PSF) for X-ray sources is about 6 arcsec.
There are 55,532 X-ray sources in the HSC--Wide region.
According to the user's guide\footnote{\url{http://xmmssc.irap.omp.eu/Catalogue/4XMM-DR11/4XMM-DR11_Catalogue_User_Guide.html}}, we make a clean point source sample with {\tt SC\_SUM\_FLAG} $< 3$ and {\tt SC\_EXTENT} = 0 (see also \cite{Nebot}).
We also apply a flux cut to selected objects, leaving 38,707 X-ray sources with $F_{\rm X}$ (0.2 -- 12 keV) $> 1.0 \times 10^{-14}$ erg s$^{-1}$ cm $^{-2}$.

We then cross-match the X-ray sample with CAMIRA member galaxies using a 4 arcsec search radius following \citet{Terashima}, which leaves 263 X-ray sources in CAMIRA clusters.
Among them, 25 XMM sources have more than two candidate counterparts of HSC sources.
We select the closest object as the counterpart.
We then estimate their rest-frame 2--10 keV X-ray luminosity ($L_{\rm X}$ (2--10 keV)) by assuming a photon index of $\Gamma = 2$ (e.g., \cite{Toba22})\footnote{We utilized {\tt XSPEC} version 12.11.1 (\cite{Arnaud}) for this calculation.}.
Figure \ref{z_L}(c) shows resultant $L_{\rm X}$ (2--10 keV) of 263 X-ray sources as a function of redshift.
Finally, we extract 236 sources with $L_{\rm X}$ (2--10 keV) $\geq 1.0 \times 10^{42}$ erg s$^{-1}$ as AGN in the same manner as \citet{Toba22}.
We note that the choice of the threshold X-ray luminosity for AGN does not significantly affect the conclusion of this work, which is also demonstrated by \citet{Bufanda}.

\subsubsection{Final AGN sample}
\label{S_AGN}

\begin{figure}[h]
 \begin{center}
 \includegraphics[width=0.45\textwidth]{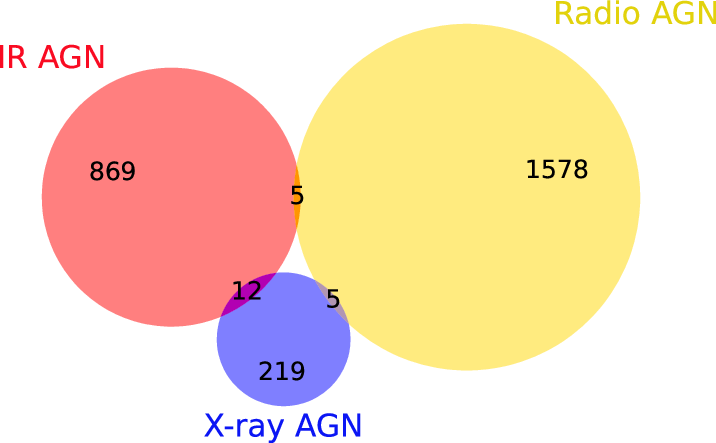}
 \end{center}
\caption{Venn diagram showing the overlap between the AGN samples constructed by IR (red), radio (yellow), and hard X-ray (blue).}
\label{venns_AGN}
\end{figure}

\begin{figure}[h]
 \begin{center}
 \includegraphics[width=0.45\textwidth]{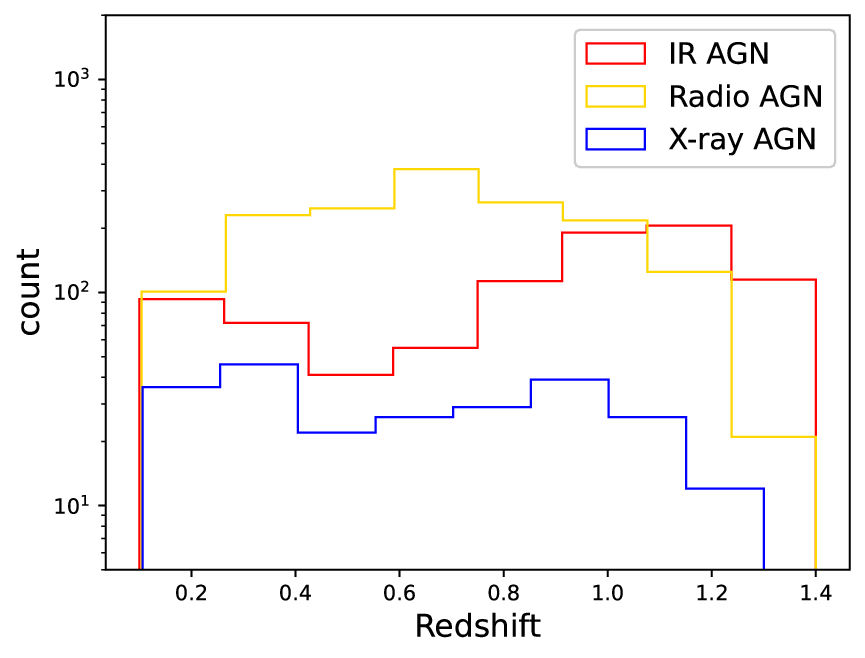}
 \end{center}
\caption{Redshift distributions of AGN selected with IR (red), radio (yellow), and hard X-ray (blue).}
\label{zhist_AGN}
\end{figure}

We identify 886, 1,588, and 236 member galaxies hosting AGN detected by MIR, Radio, and hard X-ray, respectively, as summarized in table \ref{AGN_catalog}.
Figure \ref{venns_AGN} shows the overlap between the three AGN samples.
We find that 5, 12, and 5 objects are detected by MIR/radio, MIR/X-ray, and Radio/X-ray, respectively.
No objects are detected by all wavelengths: IR, radio, and X-ray.
In summary, we select 2,688 AGN by taking into account duplication.
The redshift distributions of each AGN sample are shown in figure \ref{zhist_AGN}.

We remind readers that cross-identifications of CAMIRA member galaxies with IR, Radio, and X-ray catalogs are based on the nearest matching; we always choose the nearest one as the counterpart within the search radius.
We confirm that even if we randomly pick up an object as a counterpart from the candidates within the search radius and select AGN, it will not affect our conclusions.
We also note that the difference in spatial resolution among multi-wavelength catalogs may cause objects in especially crowded regions (groups/clusters) to be affected by blending; IR, radio, and X-ray emissions may come from multiple HSC sources in some cases.
An approach to solve this issue is an image de-blending with such as the Bayesian technique and deep learning (e.g., \cite{Merlin,Hurley,Reiman}), which is beyond the scope of this work.
Another approach is s to examine their spectral energy distributions (SEDs) to see if these are plausible, which will be addressed by Y.Toba et al. (in preparation).
We confirm that, at least for radio sources, their IR (WISE) flux seems reasonable (see section \ref{dis_Radio}).

\section{Results}
\label{Results}

\subsection{AGN fraction distribution}
\label{r_fAGN}
  
We first check how many CAMIRA clusters have member galaxies hosting AGN detected by IR, radio, and/or X-ray.
We find that 2,536 clusters have at least one AGN, as summarized in table \ref{CAMIRA_AGN}.
This means that about 9.4\% (2.536/27,037) CAMIRA clusters have AGN.

\begin{table}[h]
\tbl{CAMIRA clusters with member galaxies hosting AGN.}{
\scalebox{0.85}[0.9]{
\begin{tabular}{cr}
\hline\hline
\multicolumn{1}{c}{Wavelength}	&  \multicolumn{1}{c}{Number of clusters with member galaxies hosting AGN}	\\
\hline
IR		&	859	\\
Radio	&	1,546   \\
X-ray	&	 223    \\
\cline{2-2}
		&	Total (after removing duplication): 2,536 \\
\hline
\end{tabular}
}
}
\label{CAMIRA_AGN}
\end{table}

\begin{figure}[h]
 \begin{center}
 \includegraphics[width=0.45\textwidth]{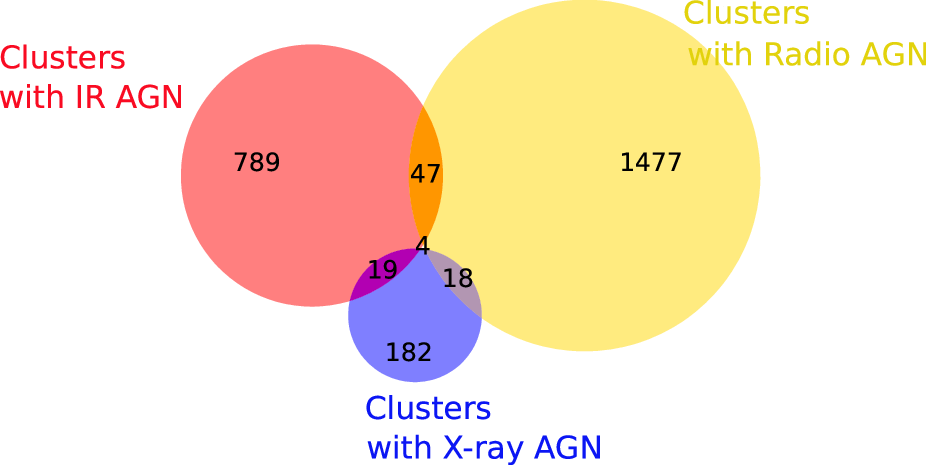}
 \end{center}
\caption{Venn diagram showing the overlap between the CAMIRA clusters in which member galaxies host AGN detected by IR (red), radio (yellow), and hard X-ray (blue).}
\label{venns_cl}
\end{figure}

\begin{figure}[h]
 \begin{center}
 \includegraphics[width=0.45\textwidth]{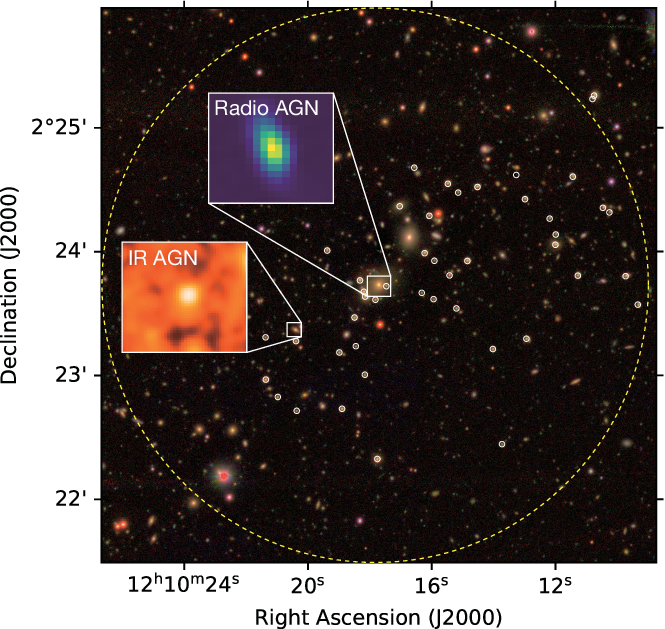}
 \end{center}
\caption{HSC color image of a CAMIRA cluster (C21a1sW14126) at $z_{\rm cl}$ = 0.386 with 2 AGN. The dotted yellow line represents a circle with a radius of $R_{\rm 200}$ ($\sim$ 1.4 Mpc). The white circles/squares indicate member galaxies of C21a1sW14126. The inserted figures show images (60\arcsec $\times$ 60\arcsec) for IR and radio AGN.}
\label{CAMIRA_image}
\end{figure}

Figure \ref{venns_cl} shows the overlap between CAMIRA clusters with member galaxies hosting AGN.
We find that 859, 1,546, and 223 galaxy clusters have AGN detected by IR, radio, and hard X-ray, respectively.
51, 23, and 22 clusters have AGN detected in MIR/radio, MIR/X-ray, and radio/X-ray, respectively.
An image of such a CAMIRA cluster with AGN detected at multiple wavelengths is shown as an example in figure \ref{CAMIRA_image}. 

\begin{figure}[h]
 \begin{center}
 \includegraphics[width=0.45\textwidth]{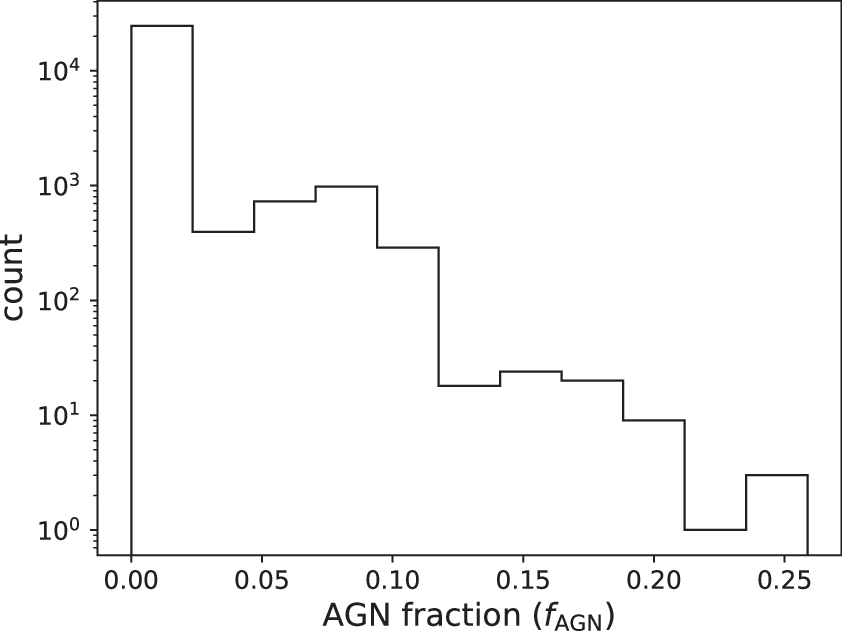}
 \end{center}
\caption{Histogram of AGN fraction ($f_{\rm AGN}$) in CAMIRA clusters.}
\label{fAGN_hist}
\end{figure}

\begin{figure}[h]
 \begin{center}
 \includegraphics[width=0.45\textwidth]{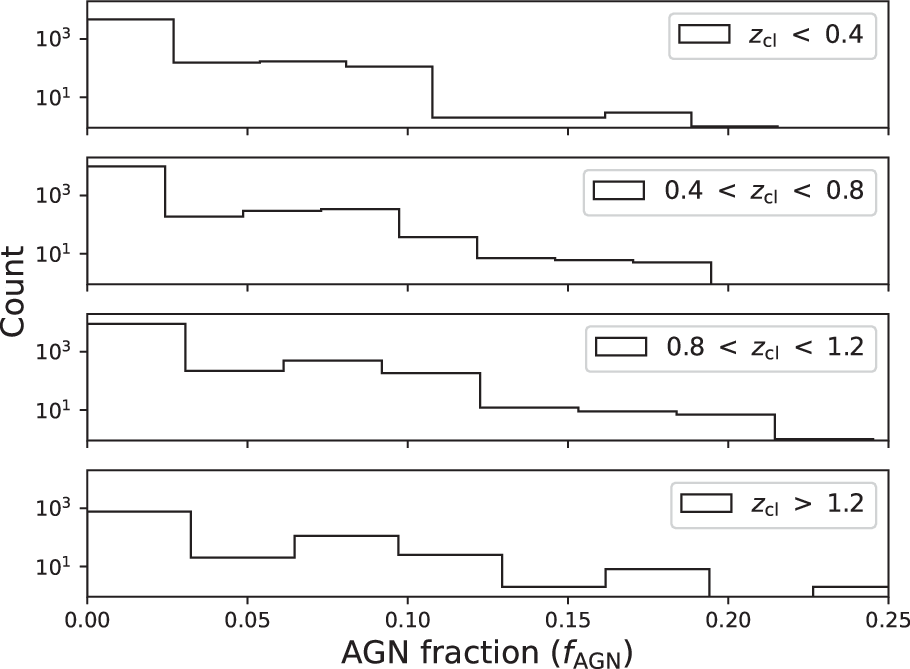}
 \end{center}
\caption{Histogram of AGN fraction ($f_{\rm AGN}$) in CAMIRA clusters for each redshift bin.}
\label{fAGN_hist_zbin}
\end{figure}

We then estimate the AGN ``number'' fraction ($f_{\rm AGN}$) in each cluster by dividing the number of AGN in a cluster by its richness ($N_{\rm mem}$).
Figure \ref{fAGN_hist} shows the distribution of $f_{\rm AGN}$ in CAMIRA clusters.
Since roughly 90\% of CAMIRA clusters do not have AGN, the histogram peaks at 0 but is distributed up to about 0.25.
We also divide the cluster sample into sub-sample based on redshifts (i.e., $z_{\rm cl} < 0.4$, $0.4 < z_{\rm cl} < 0.8$, $0.8 < z_{\rm cl} < 1.2$, and $z_{\rm cl} > 1.2$), and investigate $f_{\rm AGN}$ distribution of clusters for each redshift bin, as shown in figure \ref{fAGN_hist_zbin}.
We confirm that $f_{\rm AGN}$ is broadly distributed regardless of cluster redshift, which should be kept in mind for the following discussion.

\subsection{Redshift and cluster-centric radius dependences on AGN fraction in galaxy clusters}
\label{S_fAGN}

We present how the AGN fraction depends on redshift and distance from the cluster center ($R$, cluster-centric radius) and compare the AGN fraction with field galaxies.
One caution here is the AGN fraction in the following discussion differs from what is defined in section \ref{r_fAGN}.
To improve the statics for our sample, we combine the redshift and radial distribution of member galaxies/AGN in each bin.
In other words, we calculate a ratio of the number of AGN to the number of objects in a certain bin as an AGN fraction, which allows us to compare AGN fraction in clusters even to field galaxies (see sections \ref{S_z_fAGN} and \ref{S_dist_fAGN}).

\subsubsection{Redshift dependence on AGN fraction}
\label{S_z_fAGN}

\begin{figure}[h]
 \begin{center}
 \includegraphics[width=0.45\textwidth]{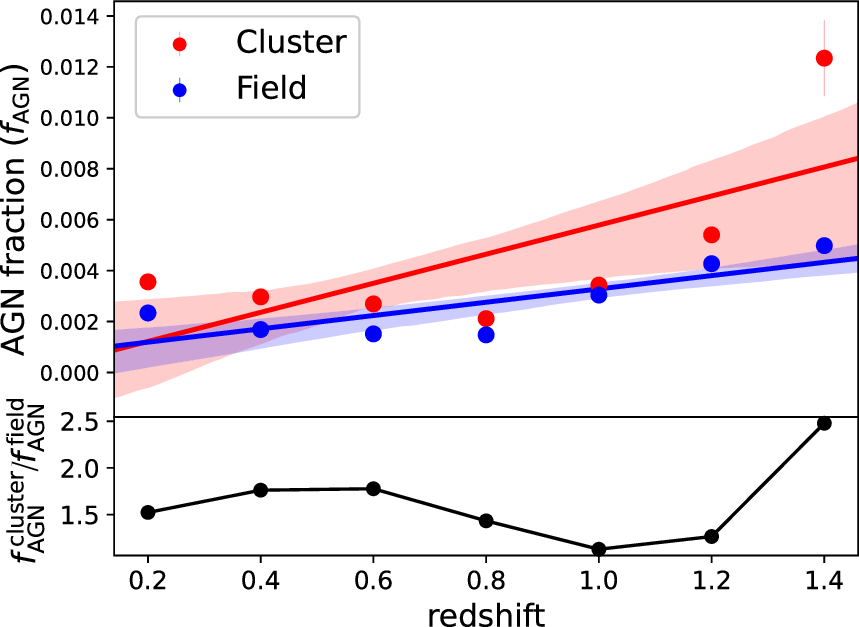}
 \end{center}
\caption{The top panel shows AGN fraction ($f_{\rm AGN}$) as a function of redshift. The red and blue circle represents $f_{\rm AGN}$ in CAMIRA clusters and field, respectively. The vertical error bars are calculated from the Poisson statistical uncertainty. The solid lines with shaded regions are best-fit linear regressions with 1$\sigma$ confidence intervals. The bottom panel shows the ratio of $f_{\rm AGN}^{\rm cluster}$ and $f_{\rm AGN}^{\rm field}$.}
\label{z_fAGN}
\end{figure}

Figure \ref{z_fAGN} shows the AGN fraction as a function of redshift.
We confirm that $f_{\rm AGN}$ increases with redshift as previous works reported (e.g., \cite{Galametz,Martini09,Haggard}).
We fit the $z_{\rm mem}$--$f_{\rm AGN}$ relation to linear regression by considering the uncertainties in each redshift bin.
We also calculate the correlation coefficient ($r_{\rm cl}$) by using the Bayesian regression method \citep{Kelly}, which provides a correlation coefficient with uncertainty (see, e.g., \cite{Toba19b,Toba21}).
The resultant value is $r_{\rm cl}$ = 0.53 $\pm$ 0.35, indicating a positive correlation between redshift and AGN fraction.
We thus confirm the existence of a Butcher-Oemler-like effect for AGN in galaxy clusters.

It should be noted that a positive correlation between redshift and $f_{\rm AGN}$ for ``field galaxies'' has also been reported by many authors (e.g., \cite{Silverman,Haggard,Oi}).
To discriminate whether the observed trend is somehow specific to AGN in galaxy clusters or we just see a trend in field galaxies, we measure the AGN fraction for field galaxies detected by the HSC-SSP.
Field galaxies (detected by the HSC) are selected by applying the same magnitude and color cuts as red sequence galaxies \citep{Oguri,Oguri18}, but not belonging to a galaxy cluster.
The redshift range of field galaxies is the same as CAMIRA member galaxies.
We identify field galaxies hosting MIR, radio, and/or X-ray AGN in the same manner as what is presented in section \ref{s_AGN}.
Figure \ref{z_fAGN} also shows $f_{\rm AGN}$ for field galaxies as a function of redshift.
We find that $f_{\rm AGN}$ for member galaxies in CAMIRA clusters is systematically higher than that for field galaxies regardless of redshift.
Typical excess (i.e., $f^{\rm cluster}_{\rm AGN}$/$f^{\rm field}_{\rm AGN}$) is about 1.6 and reaches up to 2.6 at $z > 1.2$ as shown in the bottom panel of figure \ref{z_fAGN}.
We confirm that the AGN fraction in field galaxies increases with redshift.
But we find that the best-fit slope for clusters is steeper than that for the field (see figure \ref{z_fAGN}), i.e.,  AGN in galaxies cluster increases more rapidly with redshift than the field, which is consistent with what reported in \citet{Eastman}.
Those results indicate that a denser environment (i.e., galaxy groups and clusters), particularly at high-$z$ Universe, is likely to enhance AGN activity over a wide redshift range.

\subsubsection{Cluster-centric radius dependences on AGN fraction}
\label{S_dist_fAGN}

\begin{figure}[h]
 \begin{center}
 \includegraphics[width=0.45\textwidth]{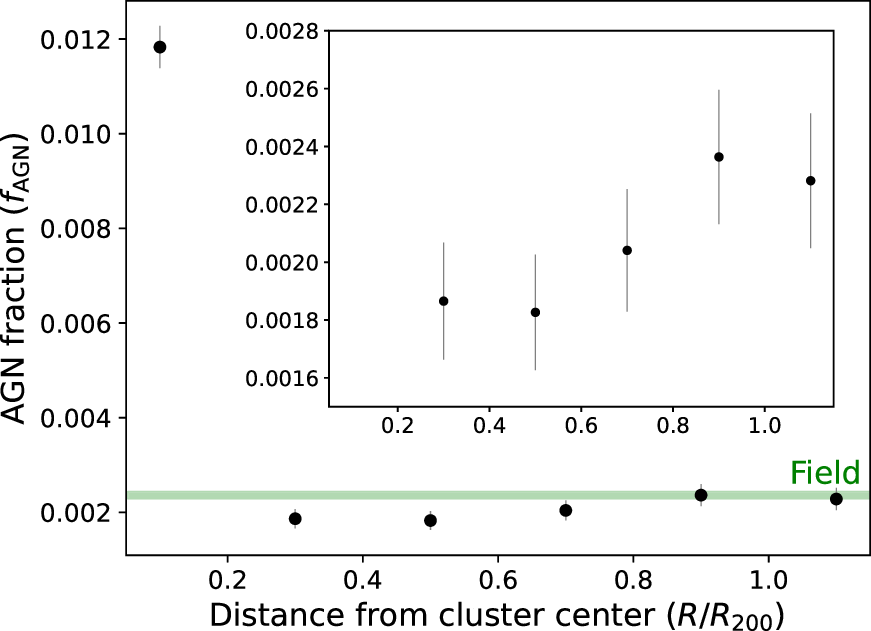}
 \end{center}
\caption{AGN fraction as a function of cluster-centric radius scaled by Virial radius ($R/R_{\rm 200}$). The vertical error bars are calculated from the Poisson statistical uncertainty. The inserted figure displays $f_{\rm AGN}$ only for $R/R_{\rm 200} > 0.1$. The green-shaded region shows an average AGN fraction in the field with uncertainty.}
\label{dist_fAGN}
\end{figure}

Figure \ref{dist_fAGN} shows the AGN fraction as a function of the projected distance from the cluster center (i.e., cluster-centric radius) scaled by Virial radius ($R_{\rm 200}$), $R/R_{\rm 200}$, where cluster centers are defined by centroids of BCGs identified by the CAMIRA algorithm \citep{Oguri18}.
$R_{\rm 200}$ is the radius within which the mass density is 200 times the mean mass density of the Universe.
We obtain $R_{\rm 200}$ by assuming a scaling relation between $N_{\rm mem}$ and cluster mass ($M_{\rm 200}$) for CAMIRA clusters reported in \citet{Okabe} (see also \cite{Murata,Chiu}), where $M_{\rm 200}$ is the total mass enclosed within a sphere of $R_{\rm 200}$.

We find a significant excess of $f_{\rm AGN}$ in the cluster center, $f_{\rm AGN}$ at $R/R_{\rm 200}$ = 0.1 is about 5 times higher than that in the outskirts of galaxy clusters and/of field galaxies, which is consistent with previous works (e.g., \cite{Mo,Li}).
This result also supports our finding in section \ref{S_z_fAGN} that AGN tends to have more emerged in denser environments.
On the other hand, $f_{\rm AGN}$ could also show a small excess toward the outskirts of galaxy clusters (see the inserted figure of figure \ref{dist_fAGN}).
Several studies indeed reported an enhancement of AGN in the outskirts of the galaxy cluster rather than the cluster center (e.g., \cite{Khabiboulline,Koulouridis}).
We will discuss this in section \ref{S_fAGN_multi}.

\section{Discussion}
\label{S_dis}

\subsection{Possible uncertainties and selection bias}

\subsubsection{Contamination of fore/background galaxies}
\label{s_contami}

In this work, we focus on member galaxies that securely belong to galaxy groups/clusters by comparing $z_{\rm cl}$ and $z_{\rm mem}$ as described in section \ref{s_member}.
Here we discuss how the possible contamination from the foreground and background galaxies could affect the redshift dependence of the AGN fraction from a membership probability point of view.
For each galaxy in the CAMIRA cluster catalog, a weight factor ($w_{\rm mem}$) corresponding to a membership probability is assigned by adopting the fast Fourier transform to the two-dimensional richness map.
A full description of $w_{\rm mem}$ is provided in \citet{Oguri}.
$w_{\rm mem}$ ranges between 0 and 1, implying that a galaxy with a value closer to 1 is more likely to belong to a galaxy cluster (i.e., high membership probability).
We examine how the correlation of $z_{\rm mem}$--$f_{\rm AGN}$ depends on membership probability.

\begin{figure}[h]
 \begin{center}
 \includegraphics[width=0.45\textwidth]{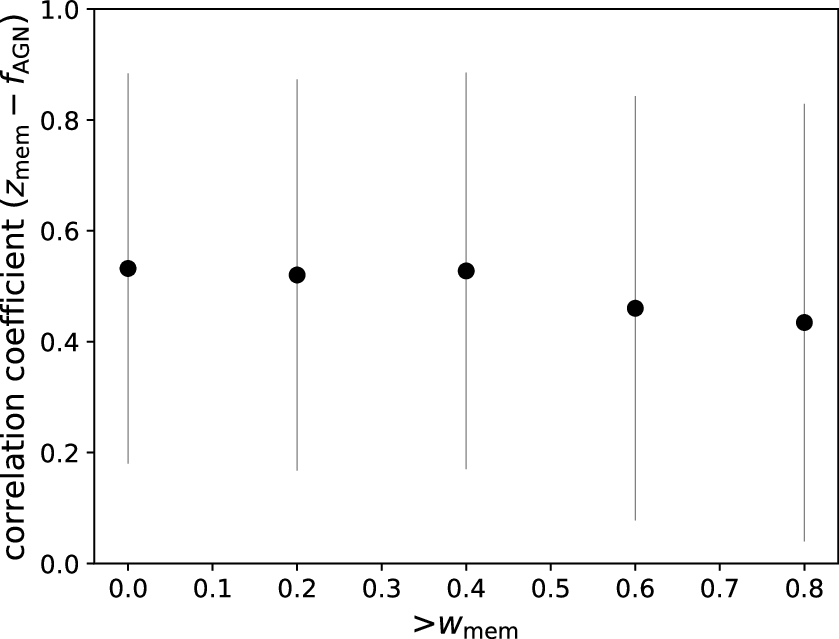}
 \end{center}
\caption{Correlation coefficients of $z_{\rm mem}$--$f_{\rm AGN}$ as a function of a weight factor $w_{\rm mem}$ (corresponding to membership probability).}
\label{z_fAGN_pmem}
\end{figure}

Figure \ref{z_fAGN_pmem} shows the resultant correlation coefficients of $z_{\rm mem}$--$f_{\rm AGN}$ as a function of $w_{\rm mem}$.
For example, the leftmost data point represents the correlation coefficient when objects with $w_{\rm mem} > 0$.0 (i.e., all member galaxy candidates) are used. 
In contrast, the rightmost data point represents the correlation coefficient when only member galaxy candidates with $w_{\rm mem} > 0.8$ are used.
We find that there is no significant $w_{\rm mem}$ dependence with a correlation coefficient (of the data points in figure \ref{z_fAGN_pmem}) being $r = -0.04$, which indicates that the effect of possible contamination from the foreground and background galaxies is negligible from a statistical point of view.

\subsubsection{Selection method for radio AGN}
\label{dis_Radio}

In section \ref{Radio_AGN}, we selected radio AGN by adopting the luminosity cut, $L_{\rm 1.4 GHz} \geq 1.0 \times 10^{24}$ W Hz$^{-1}$.
This selection practically classifies all radio-detected sources at $z_{\rm mem} > 0.5$ as radio AGN because our sample is flux-limited and so that luminosity correlates with redshift (see figure \ref{z_L}b), which might be too simple.
Some radio luminous, massive star-forming galaxies (SFGs) may be contaminated to our radio AGN sample, particularly at higher redshift if considering the star-forming main sequence (e.g., \cite{Noeske,Schreiber,Pearson}) and the IR--radio luminosity correlation of SFGs (e.g., \cite{Helou,Ivison,Delvecchio}).
We introduce another selection method for AGN based on the ratio of IR and radio luminosity ($q_{\rm IR}$), and see how the selected radio AGN sample is different from that with luminosity cut.

$q_{\rm IR}$ is defined as follows (\cite{Helou});
\begin{equation}
\label{qir_eq}
q_{\rm IR} = \log \left( \frac{L_{\rm IR} / 3.75\times 10^{12}}{L_{\rm 1.4\, GHz}} \right),
\end{equation}
where $L_{\rm IR}$ is the total IR luminosity while $3.75 \times 10^{11}$ is the frequency ($\sim$80 $\micron$) that is used for making $q_{\rm IR}$ a dimensionless quantity.
Because Far-IR (FIR) data are critical to obtaining accurate $L_{\rm IR}$, we first limit our radio sample to sources in the survey footprint of H-ATLAS DR1 (\cite{Valiante}), leaving 339 radio-detected objects. 
We then perform the SED fitting for those objects in the same manner as \citet{Toba19,Toba20} in which a SED fitting code, {\tt CIGALE} (Code Investigating GALaxy Emission; \cite{Boquien}) is employed.
The FIR flux-boosting is corrected following \citet{Toba21b}.

\begin{figure}[h]
 \begin{center}
 \includegraphics[width=0.45\textwidth]{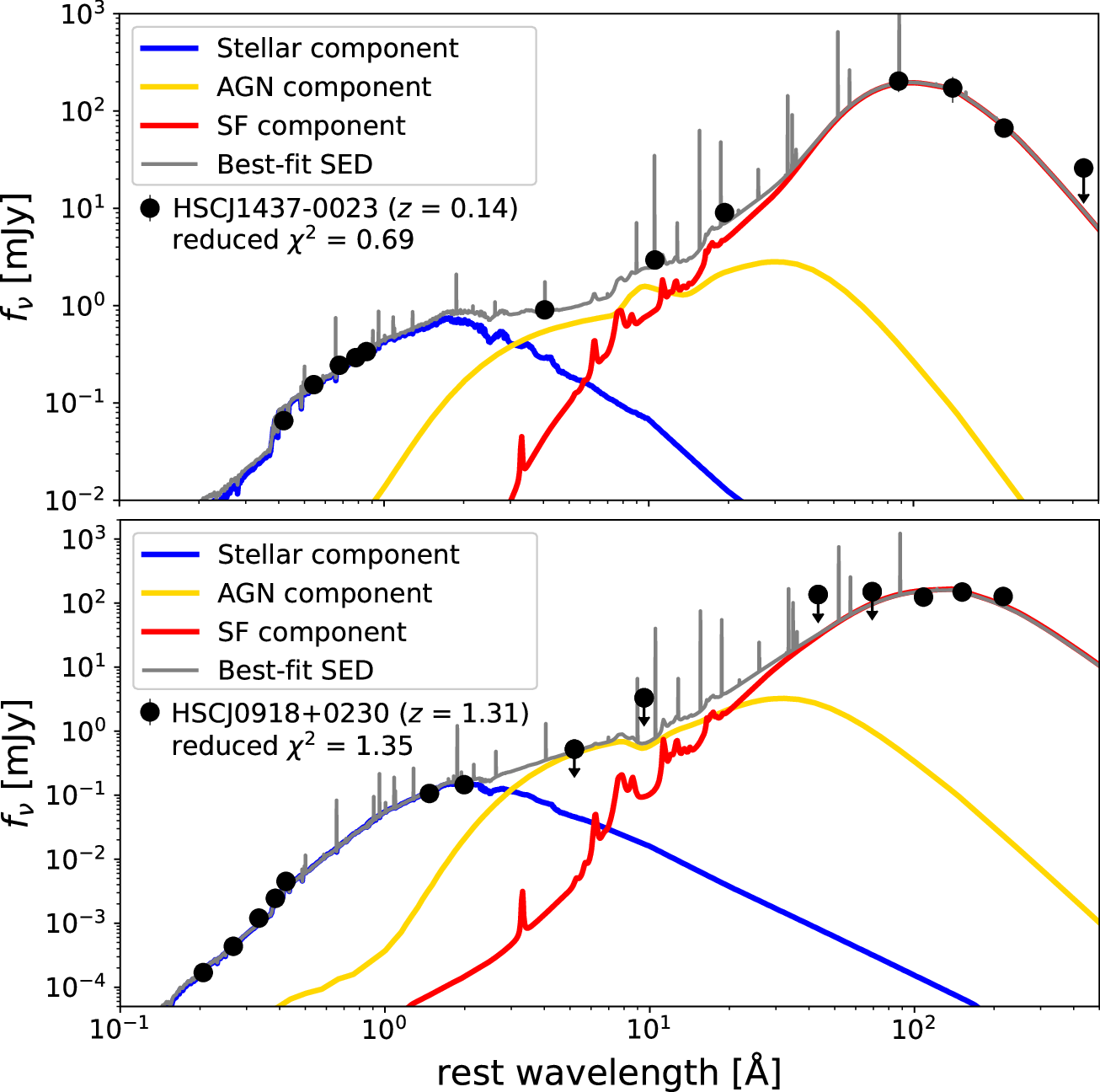}
 \end{center}
\caption{Examples of the SED fitting. The black points are photometric data. The blue, yellow, red, and green lines show stellar, AGN, SF, and radio components, respectively. The solid black lines represent the resultant SEDs.}
\label{SED_fitting}
\end{figure}

\begin{figure}[h]
 \begin{center}
 \includegraphics[width=0.45\textwidth]{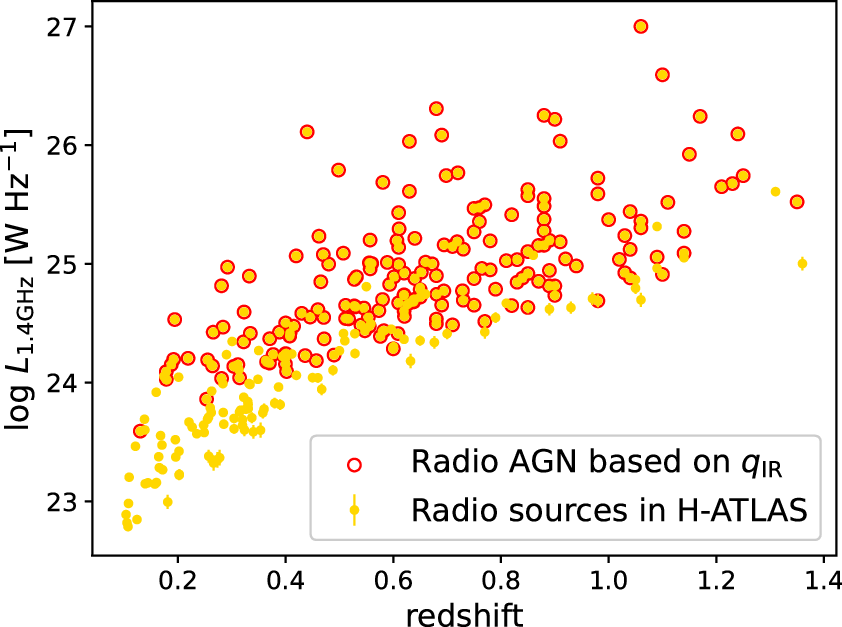}
 \end{center}
\caption{The 1.4 GHz luminosity as a function of redshift for radio sources in H-ATLAS. Red open circles denote radio AGN with $q_{\rm IR} < 2.40$.}
\label{z_Lradio_qir}
\end{figure}

Figure \ref{SED_fitting} shows examples of the SED fitting for low-$z$ and high-$z$ radio sources.
We confirm that 303/339 ($\sim$90\%) objects have reduced $\chi^2 < 1.5$, implying the data are well fitted with the combination of the stellar, AGN, and SF components by {\tt CIGALE}.
It is known that radio AGN tend to deviate from IR--radio luminosity correlation of SFGs; AGN have small $q_{\rm I}$ compared to SFGs (e.g., \cite{Sajina,Williams}).
In this work, we employ $q_{\rm IR} = 2.40$ as a threshold between SFGs and AGN \citep{Ivison}.
Figure \ref{z_Lradio_qir} shows 1.4 GHz luminosity as a function of redshift for radio-detected sources in H-ATLAS.
Radio AGN selected with $q_{\rm IR}$ cut is also over-plotted with red open circles.
We find that about 80\%\footnote{This value depends on the definition of $q_{\rm IR}$ and threshold value. If we use $L_{\rm FIR}$ instead of $L_{\rm IR}$ and employ $q_{\rm FIR} = 2.36$ as a threshold value \citep{Bell}, about 95\% of radio AGN based on radio luminosity is also classified as radio AGN.} of AGN based on radio luminosity is also classified as AGN based on $q_{\rm IR}$.
In contrast, only 2\% of SFGs with $L_{\rm 1.4\,GHz} < 10^{24}$ W Hz$^{-1}$ are classified as AGN based on $q_{\rm IR}$.
Less luminous AGN tend not to be classified as AGN through the $a_{\rm IR}$ selection, which is reasonable by definition of $q_{\rm IR}$. 
Given the consistency between the two methods, we conclude that the difference in the selection method for radio AGN does not significantly affect our results and conclusion.

\subsubsection{Stellar mass dependence}

We report a positive correlation between AGN fraction and redshift both for member galaxies in galaxy groups/clusters and field galaxies (see section \ref{S_z_fAGN}).
One caution here is that some authors reported that AGN fraction depends on the stellar mass ($M_*$) of AGN hosts (e.g., \cite{Kauffmann,Best05,Koss,Miraghaei}).
Considering that our AGN samples are flux-limited, AGN could become more pronounced for more massive host galaxies at higher redshifts, which could induce a positive correlation between $f_{\rm AGN}$ and redshift.

\begin{figure}[h]
 \begin{center}
 \includegraphics[width=0.45\textwidth]{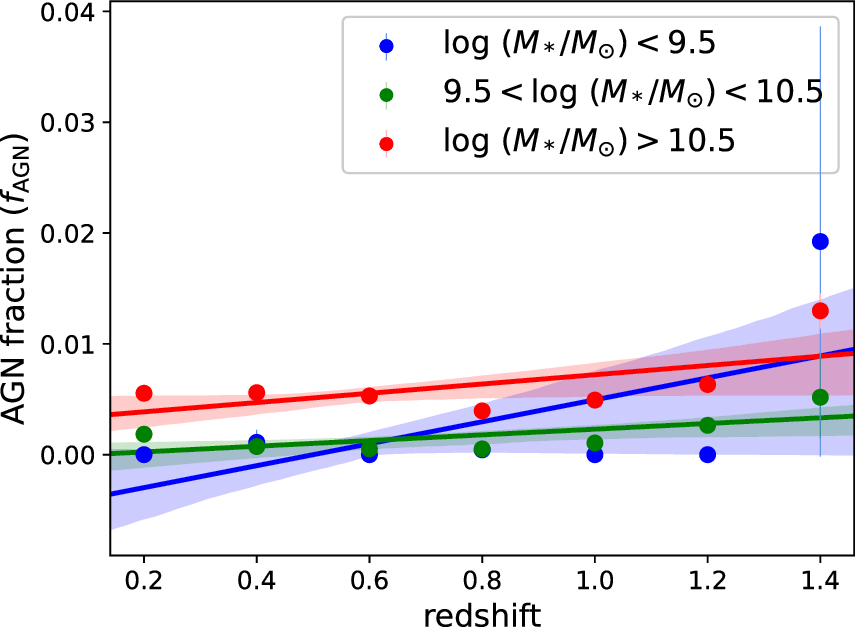}
 \end{center}
\caption{AGN fraction as a function of redshift for CAMIRA member galaxies with different stellar mass (blue: $\log\, (M_*/M_{\odot}) < 9.5$, green: $9.5 < \log\, (M_*/M_{\odot}) < 10.5$, red: $\log\, (M_*/M_{\odot}) > 10.5$). The vertical error bars are calculated from the Poisson statistical uncertainty. The solid lines with shaded regions are best-fit linear regressions with 1$\sigma$ confidence intervals.}
\label{z_fAGN_stellar}
\end{figure}

To investigate how the stellar mass of member galaxies could influence the correlation, we divide our sample into three sub-samples based on the stellar mass ($\log\, (M_*/M_{\odot}) < 9.5$, $9.5 < \log\, (M_*/M_{\odot}) < 10.5$, and $\log\, (M_*/M_{\odot}) > 10.5$), in which $M_*$ is obtained from outputs by {\tt DEmP} (see section \ref{s_member}).
Figure \ref{z_fAGN_stellar} shows $f_{\rm AGN}$ as a function of redshift for three sub-samples.
We confirm that $f_{\rm AGN}$ gets slightly larger for more massive stellar mass over the wide redshift range.
This result is consistent with \citet{Pimbblet}, who also argued the stellar mass dependence on AGN fraction in galaxy clusters and reported that AGN fraction depends on stellar mass even in the cluster environment.
On the other hand, no significant differences in correlation coefficients among the three subsamples within errors.

\subsection{Multi-wavelength view of AGN fraction}
\label{S_fAGN_multi}

We find in section \ref{S_fAGN} that the AGN number fraction increases with cluster redshift while it decreases with distance from the cluster center.
Here we discuss how the above trends depend on the AGN population (i.e., IR-, radio-, and X-ray- selected AGN).

\begin{figure}[h]
 \begin{center}
 \includegraphics[width=0.45\textwidth]{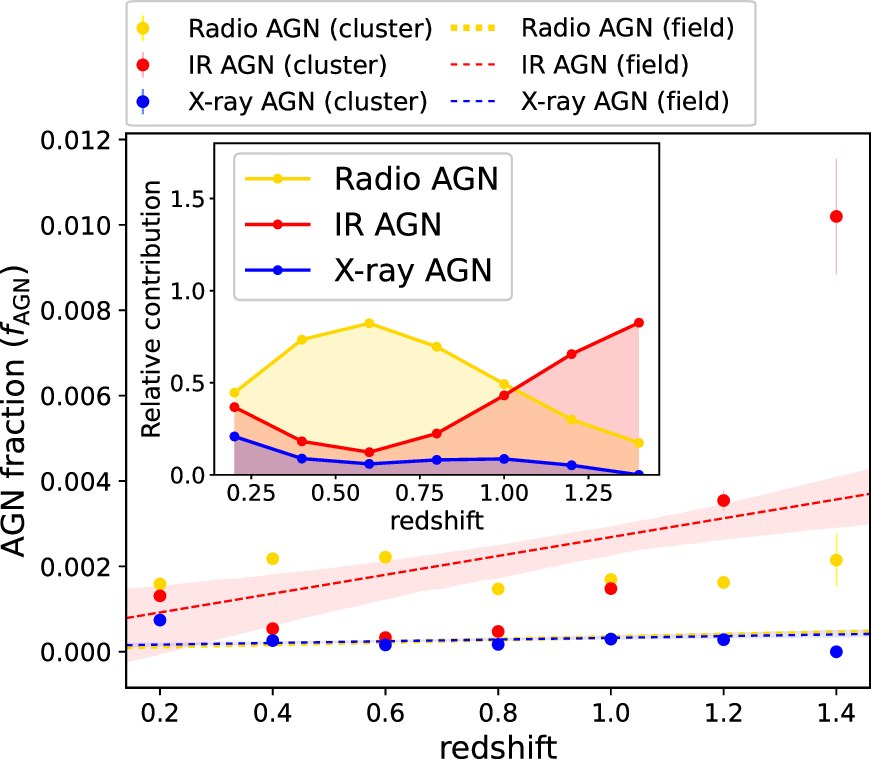}
 \end{center}
\caption{AGN fraction ($f_{\rm AGN}$) as a function of redshift for each AGN population in CAMIRA clusters. The yellow, red, and blue points denote radio-, IR-, and X-ray-detected AGN, respectively. The dashed lines denote the AGN fraction for each AGN population in the field. The inserted figure represents the relative contribution from each AGN population to $f_{\rm AGN}$.}
\label{z_fAGN_multi}
\end{figure}

Figure \ref{z_fAGN_multi} shows the AGN fraction estimated by each AGN population as a function of redshift.
The inserted figure represents the relative contribution of each AGN population to the AGN fraction, i.e., $f^{\rm IR}_{\rm  AGN}/f_{\rm AGN} + f^{\rm Radio}_{\rm  AGN}/f_{\rm AGN}  + f^{\rm X-ray}_{\rm  AGN}/f_{\rm AGN}  = 1$ for each redshift bin.
We find that the positive correlation obtained in this work is mainly due to a significant contribution from IR AGN to $f_{\rm AGN}$ at $z > 1$ (see the inserted figure in figure \ref{z_fAGN_multi}). 
This rapid increase of IR AGN at higher redshift is also reported in \citet{Tomczak}, who conducted a census of MIR-selected AGN in galaxy clusters at $0 < z_{\rm cl} < 1.3$.
They reported that $f_{\rm AGN}$ is almost constant up to $z_{\rm cl} \sim 0.8$ while is sharply increased towards $z_{\rm cl}\sim 1.3$.
One possibility for this rapid increase in IR AGN is that the emergence of IR AGN may be more sensitive to its flux/ luminosity rather than other AGN populations: AGN fraction strongly depends on IR flux/luminosity, as reported by some works (e.g., \cite{Toba15}).

Figure \ref{z_fAGN_multi} also shows $f_{\rm AGN}$ for IR-, radio-, and X-ray-detected AGN among field galaxies.
We find that radio and X-ray AGN fractions in the field do not evolve with redshift, as is the case for clusters.
We also find that the radio AGN fraction in clusters is significantly higher than in the field, regardless of redshift.
This suggests that radio AGN may prefer a denser environment, which supports previous works (e.g., \cite{Kolwa,Uchiyama}).
On the other hand, the IR AGN fraction increases with redshift, even in the field.
This implies that the main driver for increasing $f_{\rm AGN}$ is IR AGN, independent of the environment.
But at least for $z > 1.2$, IR AGN seems to be much more abundant in the cluster as compared to the field.

\begin{figure}[h]
 \begin{center}
 \includegraphics[width=0.45\textwidth]{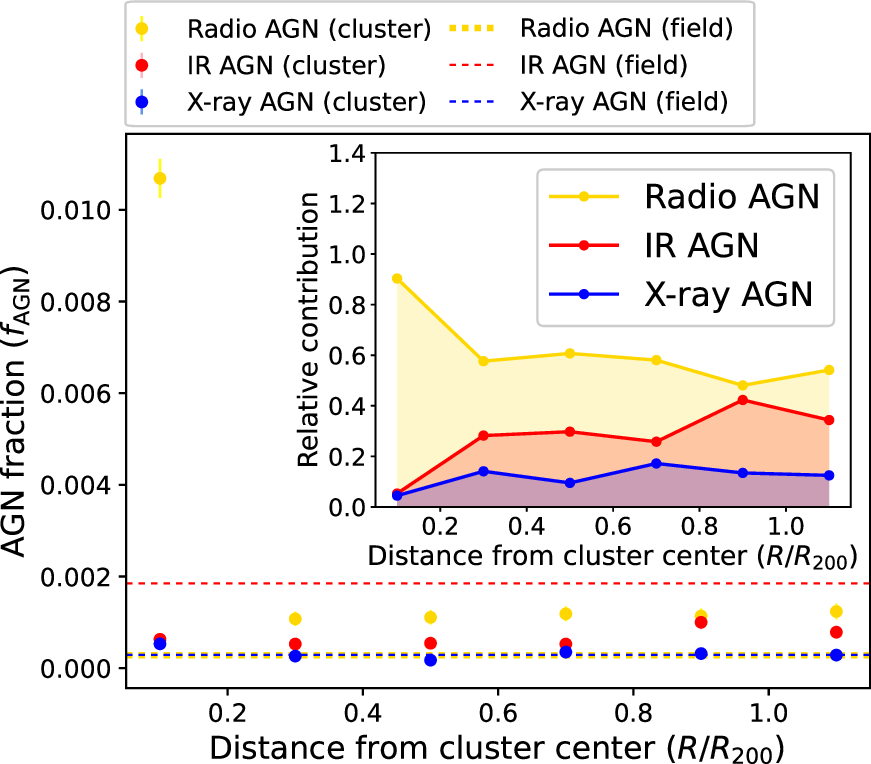}
 \end{center}
\caption{AGN fraction ($f_{\rm AGN}$) as a function of cluster-centric radius scaled by Virial radius ($R/R_{\rm 200}$). The yellow, red, and blue points denote radio-, IR-, and X-ray-detected AGN, respectively. The dashed lines denote the AGN fraction for each AGN population in the field. The inserted figure represents the relative contribution from each AGN population to $f_{\rm AGN}$.}
\label{dist_fAGN_multi}
\end{figure}

Figure \ref{dist_fAGN_multi} shows the AGN fraction for each AGN population as a function of cluster-centric radius.
The inserted figure represents the relative contribution of each AGN population to the AGN fraction for each distance bin.
We find that the excess of $f_{\rm AGN}$ in the cluster center reported in section \ref{S_dist_fAGN} is produced by radio AGN, which is in good agreement with previous works (e.g., \cite{Lin07,Mo}).
\citet{Lin17} reported that a fraction of powerful radio AGN increases with increasing stellar mass based on CAMIRA clusters.
The mean radio luminosity and stellar mass of our radio AGN in the innermost region of galaxy clusters are $\log\, L_{\rm 1.4 GHz}$ = 24.8 W Hz$^{-1}$ and $\log\, (M_*/M_{\odot}) = 11.3$ that are higher than those at the outskirts of galaxy clusters, which indicates that massive galaxies hosting powerful radio AGN could contribute to the central excess of $f_{\rm AGN}$.
In addition, since the BCGs in the cluster center tend to associate with radio AGN (e.g., \cite{Best07})\footnote{\citet{Nishizawa18} also reported that red. passive galaxies are more concentrated toward the cluster center.}, this concentrated distribution of radio AGN in galaxy clusters is reasonable. 

We also find that IR AGN may contribute to $f_{\rm AGN}$ at the outskirts of galaxy clusters rather than cluster center while the X-ray AGN fraction is almost constant up to $R/R_{\rm 200}$ = 1.0.
\citet{Klesman14} investigated the radial distribution of multiple AGN populations (e.g., IR and X-ray AGN) in galaxy clusters at $0.5 < z_{\rm cl} < 0.9$ and found that MIR AGN are less centrally concentrated compared to other AGN populations (see also \cite{Klesman12}), which supports our result.
Recently, \citet{Rodr} investigated a fraction of AGN with strong ionized gas outflow as a function of cluster-centric radius and found that that fraction could reach a peak at $\sim 0.8 R/R_{\rm 180}$ and decline towards the cluster center.
Since WISE-detected IR AGN often show strong ionized gas outflow (e.g., \cite{Toba17c}), those IR AGN with ionized gas outflow could contribute to $f_{\rm AGN}$ at the outskirts.
Another possible explanation of MIR-AGN excess in the outskirts of galaxy clusters is cluster-cluster mergers (see section \ref{S_fAGN_morph}).
We note that X-ray AGN might also show an excess of $f_{\rm AGN}$ towards the outskirts, as \citet{Koulouridis} reported (but see e.g., \cite{Montero} reporting no significant environmental dependence for X-ray AGN).

Figure \ref{dist_fAGN_multi} also shows $f_{\rm AGN}$ in the field for IR-, radio-, and X-ray-detected AGN.
In comparison to CAMIRA clusters, the radio and X-ray AGN fractions in the field are comparable or smaller than those in clusters, which is consistent with what is reported in section \ref{S_dist_fAGN}.
On the other hand, the IR AGN fraction in the field is about 1.8 times higher than that in clusters.
One possibility of this discrepancy may be the contamination of SFGs to IR AGN sample in the field (see \cite{Assef}).
Actually, the fraction of IR AGN with $\log \,(L_{\rm IR}/L_{\odot}) < 11$ in the filed sample is about 1.5 times larger than that in clusters.
Those less luminous IR galaxies do not always show a significant IR AGN signature, and IR emission from star formation might be dominant (e.g., \cite{Imanishi,Toba15}).
Hence, $f_{\rm AGN}$ for IR AGN in the field may be overestimated.
A SED-based approach can pick up even weak IR AGN and test this possibility, which will be addressed in Y.Toba et al. (in preparation).

\subsection{Cluster mass dependence on AGN fraction}
\label{S_fAGN_nass}

\begin{figure}[h]
 \begin{center}
 \includegraphics[width=0.45\textwidth]{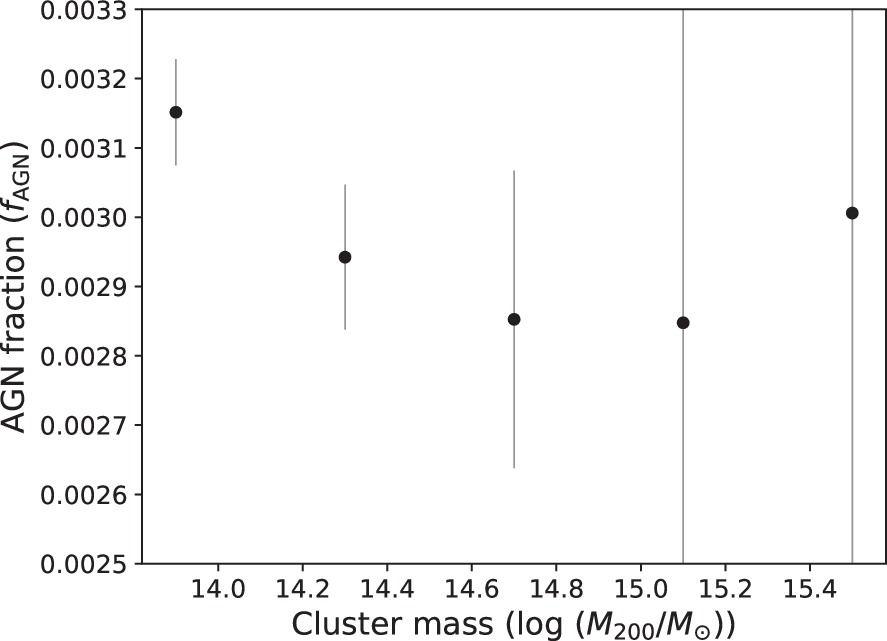}
 \end{center}
\caption{AGN fraction as a function of cluster mass ($M_{\rm 200}$).}
\label{Mcl_fAGN}
\end{figure}

We then investigate how the AGN fraction depends on cluster mass (or richness).
Figure \ref{Mcl_fAGN} shows the AGN fraction as a function of cluster mass ($M_{\rm 200}$).
We find that the AGN fraction decreases with increasing cluster mass with a correlation coefficient of $r \sim -0.2$.
This is consistent with previous works (e.g., \cite{Koulouridis18,Noordeh}).
\citet{Popesso} also reported that AGN fraction shows anti-correlation with velocity dispersion of galaxy clusters. 
Since the velocity dispersion of galaxy clusters can be translated into cluster mass (e.g., \cite{Smith}), the above work also supports our result.
Those results suggest that a galaxy in a group environment tends to ignite AGN compared to a cluster environment, with an agreement with e.g., \citet{Li,Pentericci}.

\subsection{Cluster morphology dependence on AGN fraction}
\label{S_fAGN_morph}

Finally, we discuss how the morphology of galaxy clusters would depend on the emergence of AGN.
\citet{Okabe} identified merging cluster candidates from the CAMIRA cluster catalog by using a peak-finding method.
This peak-finding method essentially counts the number of peaks above the redshift-dependent threshold based on Gaussian smoothed maps of the number densities of member galaxies.
We refer the reader to \citet{Okabe} for the full details on this method.
This work considers a galaxy cluster with a single peak as a ``relaxed'' cluster, while a galaxy cluster with more than two peaks is a ``merging'' cluster.
Consequently, 2,558/27,037 ($\sim$9.5\%) CAMIRA clusters are classified as merging clusters in which 150,684 member galaxies are associated.

\begin{figure}[h]
 \begin{center}
 \includegraphics[width=0.45\textwidth]{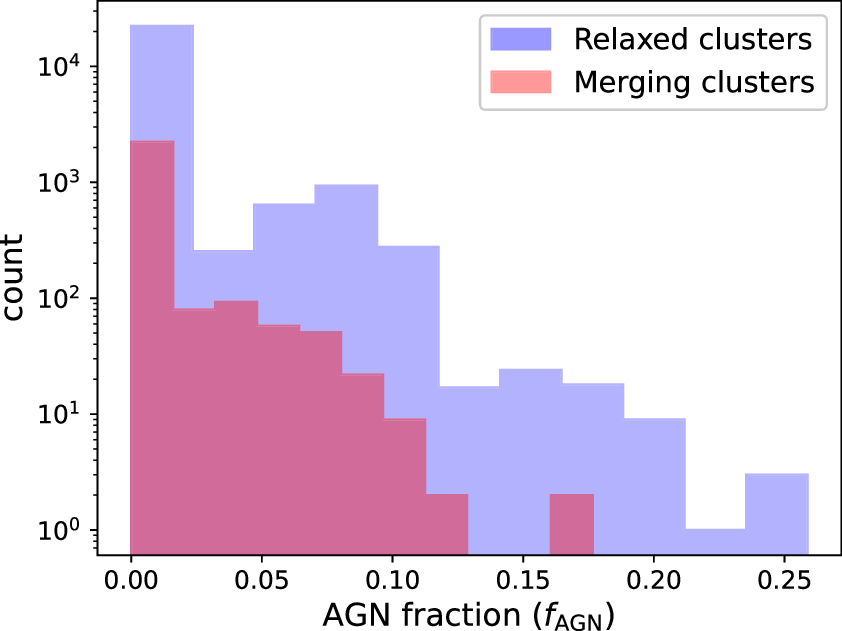}
 \end{center}
\caption{$f_{\rm AGN}$ distributions for relaxed (blue) and merging (red) clusters.}
\label{fAGN_morph}
\end{figure}

Figure \ref{fAGN_morph} shows the distribution of $f_{\rm AGN}$ for relaxed and merging clusters.
The distributions of relaxed and merging clusters are clearly different; merging clusters tend to have small $f_{\rm AGN}$, which is confirmed by a two-sided Kolmogorov-Smirnov (KS) test with $>$99.9\% significance.
This result indicates that cluster-cluster mergers may not necessarily trigger AGN.
It was suggested that cluster dynamical activity could also activate SF activity in member galaxies (e.g., \cite{Miller,Sobral,Stroe15,Okabe}).
Recently, \citet{Stroe} also reported that a large fraction of emission line galaxies in merging clusters is powered by star formation rather than AGN.
On the other hand, \citet{Noordeh} suggested that merging cluster environments might contribute to enhancing AGN activity\footnote{\citet{Bilton} also reported that merging clusters might hold relatively younger AGN populations compared with those in relaxed clusters. Further investigations based on SED fitting will test the above possibility, which will be included in future work (Y.Toba et al. in preparation).}.
Several enhancement mechanisms for SF and AGN activity in merging clusters are suggested, such as gas perturbations driven by ram-pressure and galaxy-galaxy interactions (\cite{Treu}, and reference therein).
The relative strength of SF and AGN could depend on the dominant mechanism described above and a sequence of cluster-cluster mergers.
Future statistical work, taking into account the merger stage of a cluster-cluster merger, may give us an avenue to solve this issue.

\begin{figure}[h]
 \begin{center}
 \includegraphics[width=0.45\textwidth]{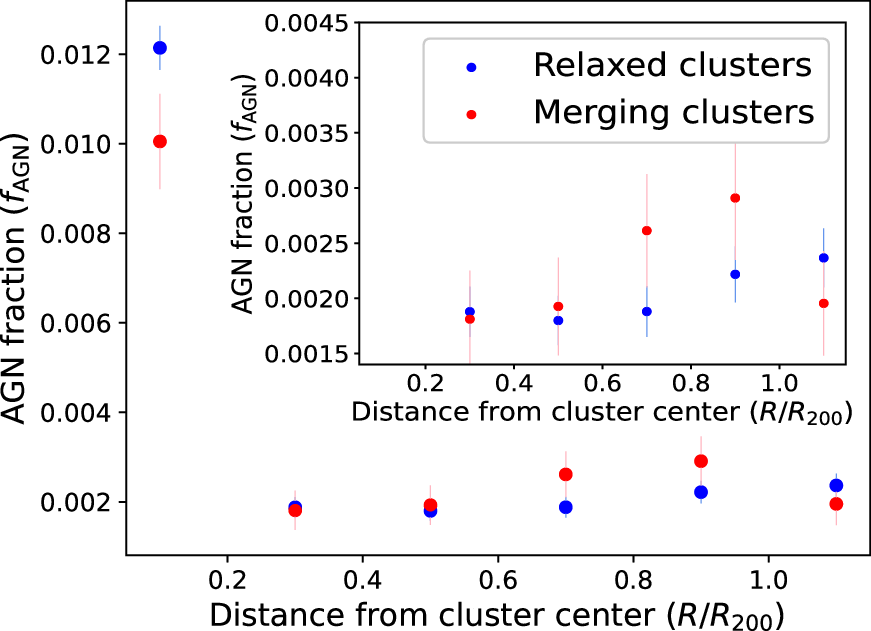}
 \end{center}
\caption{AGN fraction as a function of cluster-centric radius scaled by Virial radius ($R/R_{\rm 200}$) for relaxed (blue) and merging (red) clusters. The inserted figure displays $f_{\rm AGN}$ only for $R/R_{\rm 200} > 0.1$.}
\label{dist_fAGN_morph}
\end{figure}

Figure \ref{dist_fAGN_morph} shows cluster-centric radius dependence of AGN fraction for relaxed and merging clusters.
At the galaxy center, we find that AGN could be more enhanced in the relaxed cluster, i.e., cluster-cluster mergers are not associated with AGN enhancement in the cluster center.
The member galaxies may be too fast to interact with each other, especially in the cluster center, even if a cluster-cluster merger occurred.
With a relatively large uncertainty in mind, we also find AGN would be more enhanced at the outskirts of the emerging clusters rather than the relaxed cluster.
Given the fact that IR AGN could dominate at the outskirts of clusters (section \ref{S_fAGN_multi}), cluster-cluster mergers would enhance IR AGN at the outskirts of the clusters.
It should be noted that the cluster's optical center is not always identical to the galaxy density peak and even shows a significant offset with respect to the X-ray center in some cases (e.g., \cite{Mahdavi,Oguri18,Ota23}).
This effect could get severe for merging clusters.
Although the fraction of merging clusters is small, and thus, this effect does not significantly affect the overall trend discussed in section \ref{S_dist_fAGN}, we should keep in mind this potential uncertainty of $R$ for merging clusters.

\begin{figure}[h]
 \begin{center}
 \includegraphics[width=0.45\textwidth]{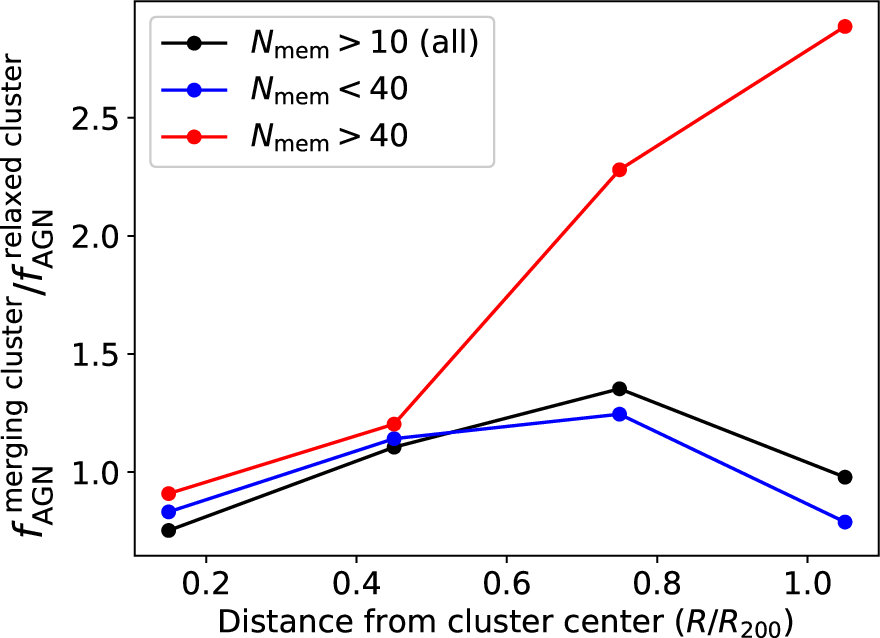}
 \end{center}
\caption{The excess of AGN fraction of merging clusters with respect to relaxed clusters, $f^{\rm merging ~ cluster}_{\rm AGN}/f^{\rm relaxed ~ cluster}_{\rm AGN}$ as a function of cluster-centric radius scaled by Virial radius ($R/R_{\rm 200}$). The blue and red lines correspond to clusters with $N_{\rm mem} < 40$ and $> 40$, respectively. The black line corresponds to clusters with $N_{\rm mem} > 10$, i.e., all cluster samples.}
\label{dist_fAGN_morph_richness}
\end{figure}

We also note that the abundance of relaxed/merging clusters may depend on richness and classification method.
Indeed, the majority of CAMIRA clusters with $N_{\rm mem} > 40$ are classified as merging clusters if we employ other classification methods such as BCG-X-ray peak offset or the concentration parameter of the X-ray surface brightness \citep{Ota23} (see also \cite{Ota20}).
This could suggest that enhancement of $f_{\rm AGN}$ for merging clusters seen in figure \ref{dist_fAGN_morph} depends on richness.
To address this issue, we divide the cluster sample with a richness of 40 as the threshold value and examine how the excess of $f_{\rm AGN}$ of merging clusters with respect to relaxed clusters in the cluster outskirts depends on richness.
Figure \ref{dist_fAGN_morph_richness} shows the excess of AGN fraction for merging clusters, $f^{\rm merging ~ cluster}_{\rm AGN}/f^{\rm relaxed ~ cluster}_{\rm AGN}$ as a function of cluster-centric radius. 
We find a significant excess of $f_{\rm AGN}$ for merging clusters with $N_{\rm mem} > 40$ in the outskirts of clusters. 
This result indicates that enhancement of AGN activity in the outskirts of clusters is preferentially occurred in merging clusters with high richness.

\section{Summary}
\label{S_summary}

We examine how AGN activity depends on the environment, particularly cluster redshift and distance from the cluster center, using the Subaru HSC-selected galaxy clusters with the CAMIRA algorithm.
The CAMIRA catalog contains more than 27,000 galaxy groups and clusters at $0.1 < z_{\rm cl} < 1.4$ with more than 1 million member galaxies, enabling us to address the above issues statistically.
We construct a multi-wavelength (IR, radio, and X-ray) selected AGN sample to identify member galaxies hosting AGN.
We find that 2,536 CAMIRA clusters have at least one AGN in their member galaxies, and 2,688 member galaxies host AGN.
Our main findings are as follows:
\begin{itemize}
\item In agreement with recent studies, AGN fraction increases with redshift for both cluster member and field galaxies.
But regardless of redshift, the AGN fraction in galaxy clusters is always higher than that in field galaxies.
In particular, a rapid increase in AGN fraction at $z_{\rm cl} > 0.8$ is caused by a significant contribution from IR-selected AGN.

\item AGN fraction increases towards cluster center. 
This is mainly because BCGs with radio AGN dominate the cluster center.
We also find a small excess of AGN fraction at the outskirts of galaxy clusters, which is mainly contributed by IR-selected AGN.

\item With the caveat of possible uncertainty of cluster centers for merging clusters in mind, cluster-cluster mergers may not be the main driver of AGN in member galaxies, especially at the cluster center.
But a cluster-cluster merger could enhance IR-AGN in the outskirts of (particularly massive) galaxy clusters.

\end{itemize}

Those results indicate that the emergence of the AGN population depends on the environment and redshift, and galaxy groups and clusters at high-$z$ play an important role in AGN evolution. 
On the other hand, a majority of member galaxies in CAMIRA clusters have not yet been spectroscopically confirmed.
Next-generation multi-object spectrographs, such as Subaru Prime Focus Spectrograph (PFS: \cite{Takada}), will overcome this issue and provide a solid conclusion.

\begin{ack}
We gratefully thank the anonymous referee for a careful reading of the manuscript and very helpful comments. 
We deeply thank Prof. John D. Silverman, Prof. Masayuki Akiyama, Dr. Connor Bottrell, and Dr. I-Non Chiu for fruitful discussion and comments.
We also thank Nari Suzuki, Manami Furuse, and Yurika Matsuo for their support.
This work is supported by JSPS KAKENHI Grant numbers JP18J01050, JP19K14759, and JP22H01266 (YT), JP20K04027 (NO), JP21K03632 (MI) and JP22K20391 and JP23K13154 (SY), and JP23K03460, JP21H05449, and JP20K22360 (TO).

The Hyper Suprime-Cam (HSC) collaboration includes the astronomical communities of Japan and Taiwan, and Princeton University. The HSC instrumentation and software were developed by the National Astronomical Observatory of Japan (NAOJ), the Kavli Institute for the Physics and Mathematics of the Universe (Kavli IPMU), the University of Tokyo, the High Energy Accelerator Research Organization (KEK), the Academia Sinica Institute for Astronomy and Astrophysics in Taiwan (ASIAA), and Princeton University. Funding was contributed by the FIRST program from the Japanese Cabinet Office, the Ministry of Education, Culture, Sports, Science and Technology (MEXT), the Japan Society for the Promotion of Science (JSPS), Japan Science and Technology Agency (JST), the Toray Science Foundation, NAOJ, Kavli IPMU, KEK, ASIAA, and Princeton University. 

This paper makes use of software developed for the Large Synoptic Survey Telescope. We thank the LSST Project for making their code available as free software at  http://dm.lsst.org

This paper is based on data collected at the Subaru Telescope and retrieved from the HSC data archive system, which is operated by the Subaru Telescope and Astronomy Data Center (ADC) at National Astronomical Observatory of Japan. Data analysis was in part carried out with the cooperation of Center for Computational Astrophysics (CfCA), National Astronomical Observatory of Japan. The Subaru Telescope is honored and grateful for the opportunity of observing the Universe from Maunakea, which has the cultural, historical and natural significance in Hawaii. 

The Pan-STARRS1 Surveys (PS1) and the PS1 public science archive have been made possible through contributions by the Institute for Astronomy, the University of Hawaii, the Pan-STARRS Project Office, the Max Planck Society and its participating institutes, the Max Planck Institute for Astronomy, Heidelberg, and the Max Planck Institute for Extraterrestrial Physics, Garching, The Johns Hopkins University, Durham University, the University of Edinburgh, the Queen’s University Belfast, the Harvard-Smithsonian Center for Astrophysics, the Las Cumbres Observatory Global Telescope Network Incorporated, the National Central University of Taiwan, the Space Telescope Science Institute, the National Aeronautics and Space Administration under grant No. NNX08AR22G issued through the Planetary Science Division of the NASA Science Mission Directorate, the National Science Foundation grant No. AST-1238877, the University of Maryland, Eotvos Lorand University (ELTE), the Los Alamos National Laboratory, and the Gordon and Betty Moore Foundation.

This publication makes use of data products from the Wide-field Infrared Survey Explorer, which is a joint project of the University of California, Los Angeles, and the Jet Propulsion Laboratory/California Institute of Technology, funded by the National Aeronautics and Space Administration.

The National Radio Astronomy Observatory is a facility of the National Science Foundation operated under cooperative agreement by Associated Universities, Inc.

This research has made use of data obtained from the 4XMM XMM-Newton serendipitous source catalogue compiled by the 10 institutes of the XMM-Newton Survey Science Centre selected by ESA.

\end{ack}


\bibliographystyle{pasj}
\bibliography{ref.bib}

\end{document}